\documentclass[prc,superscriptaddress,showpacs,twocolumn,amssymb,amsmath,amsfonts,aps]{revtex4}
\usepackage[pdftex]{graphicx} 
\usepackage{dcolumn}  
\usepackage{longtable}
\usepackage{bm}       
\usepackage{subfigure}
\usepackage{multirow} %
\begin{document}

\newcommand*{\CMU}{Carnegie Mellon University, Pittsburgh, Pennsylvania 15213, USA}
\newcommand*{\CMUindex}{1}
\affiliation{\CMU}
\newcommand*{\WJ}{Washington \& Jefferson College, Washington, Pennsylvania 15301, USA}
\newcommand*{\WJindex}{2}
\affiliation{\WJ}
\newcommand*{\GLASGOW}{University of Glasgow, Glasgow G12 8QQ, United Kingdom}
\newcommand*{\GLASGOWindex}{3}

%
%
\author {Biplab~Dey}
\affiliation{\CMU}
\author{Michael~E.~McCracken}
\affiliation{\CMU}
\affiliation{\WJ}
\author {David~G.~Ireland}
\affiliation{\GLASGOW}
\author {Curtis~A.~Meyer}
\affiliation{\CMU}

\date{\today}
\title{Polarization observables in the longitudinal basis for pseudo-scalar meson photoproduction using a density matrix approach}

\begin{abstract}
The complete expression for the intensity in pseudo-scalar meson photoproduction with a polarized beam, target, and recoil baryon is derived using a density matrix approach that offers great economy of notation. A Cartesian basis with spins for all particles quantized along a single direction, the longitudinal beam direction, is used for consistency and clarity in interpretation. A single spin-quantization axis for all particles enables the amplitudes to be written in a manifestly covariant fashion with simple relations to those of the well-known CGLN formalism. Possible sign discrepancies between theoretical amplitude-level expressions and experimentally measurable intensity profiles are dealt with carefully. Our motivation is to provide a coherent framework for coupled-channel partial-wave analysis of several meson photoproduction reactions, incorporating recently published and forthcoming polarization data from Jefferson Lab. 
\end{abstract}

\maketitle

\section{Introduction}

The production amplitude for pseudo-scalar meson photoproduction involves eight complex amplitudes which depend on the spin states of the photon and target and final-state baryons.
The reaction can be simplified by considering the parity invariance of the strong and electromagnetic interactions, reducing the number of independent complex amplitudes to four.
Barker, Donnachie, and Storrow (BDS)~\cite{bds} showed that there exist fifteen experimentally observable single- and double-polarization observables which, in addition to the differential cross-section, can be expressed as bilinears in the four independent amplitudes.
Several ambiguities originate with the BDS work. 
The BDS article treats reactions with the four amplitudes in the helicity basis (non-flip, $N$, double-flip, $D$, and two single-flip amplitudes, $S_1$ and $S_2$), but does not clearly specify to which helicity configurations the amplitudes $S_1$ and $S_2$ refer. Other authors~\cite{walker, fts} follow different schemes for enumerating the four amplitudes. 
The construction of the helicity amplitudes presents a separate problem, as choice of phase conventions in Wigner rotation matrices can also potentially lead to sign ambiguities. 

Chiang and Tabakin (CT)~\cite{ct} later showed that to completely characterize the full production amplitude, measurements of the differential cross-section and a carefully chosen set of only seven polarization observables is required; that is, there is redundancy in the full set of sixteen bilinear observables. 
The CT study assumes ``measurements'' are made with infinite precision, a situation obviously unattainable by any experiment.  More recently, the effects of uncertainty in polarization measurements on constraining amplitudes has been studied from an information theory perspective~\cite{ireland}.

Sandorfi \textit{et al.}~\cite{sandorfi} have pursued descriptions of the polarization observables in terms of the Chew-Goldberger-Low-Nambu (CGLN) amplitudes~\cite{cgln}.  They have tested configurations of production amplitudes which reproduce available polarization data for the reaction ${\gamma p \rightarrow K^{+}\Lambda}$ to within experimental uncertainties by randomly sampling the amplitude space and projecting observables from the amplitudes.  Their work shows that the currently observed set of polarization observables and experimental uncertainties do not provide enough constraint to distinguish between production models containing different resonance contributions, thus suggesting that measurement of a larger number of observables than prescribed by CT will be required to fully extract the four complex amplitudes.

The scope of the present work is three-fold. First, we derive the general expression for the reaction intensity with all three polarizations (beam, target, and recoil). 
Our motivation is the density matrix approach of Fasano, Tabakin and Saghai (FTS)~\cite{fts}, the power of which is compactness of notation. 
The full expression consists of $4 \times 4 \times 4 = 64$ terms. Invariance under mirror symmetry transformations (a parity inversion followed by a rotation, see Sec.~\ref{sec:mirror_symmetry}) removes half of these terms. In the remaining terms, each of the sixteen physically measurable observables occurs twice. 
All results herein follow simply from the properties of the Pauli matrices and the mirror symmetry operator acting on the spin density matrices of the photon and baryons. 

Second, we provide amplitude-level expressions for the polarization observables corresponding to measurable particle momentum distributions, carefully keeping track of the relative signs between experimental measurements and amplitude-level expressions. Our amplitudes are constructed in the longitudinal basis, that is, with spin projections for {\em all} particles quantized along a single direction, the beam direction. For reactions with multiple decays and non-zero spins for the the final-state particles, a single spin-quantization axis enables one to write the full production amplitude in a manifestly Lorentz-invariant fashion. A relevant example of such a process is the reaction chain $\gamma p \to K^+ \Sigma^0 \to K^+ \Lambda (\gamma) \to K^+ p \pi (\gamma)$. 

Finally, we list and numerically validate the various ``consistency relations'' connecting the different spin observables.  These consistency relations provide important checks for both theoretical analyses and constraints in the case of future experiments which will have access to polarizations of the beam and target and recoil baryons.

We aim to establish a consistent partial-wave-analysis formalism for recent~\cite{bradford-dcs, bradford, prc_klam, prc_ksig,prc_eta} and future~\cite{g8, frost} meson photoproduction data from CLAS, as well as all presently available polarization data from other experiments such as GRAAL~\cite{lleres1,lleres2} and LEPS~\cite{hicks1,sumihama,zegers}.  
To ensure correct interpretation of these observables, one must be confident of the relationship between measurements extracted from momentum distribution asymmetries and the polarization observables as treated in theoretical studies.
Until now, most of the world data on polarizations has been limited by statistics and kinematic coverage. Furthermore, experimental limitations have restricted measurement to only a few of the fifteen polarization observables.  However, with a new generation of experiments at Jefferson Lab on the horizon~\cite{g8, frost}, much activity is anticipated in the field in the coming years. This article will provide a self-contained and comprehensive description of the formalism for pseudo-scalar meson photoproduction from the fundamental derivations and a careful treatment of the connection between theory and measurement of polarization observables.

\section{Axis conventions}

\begin{figure}
\centering
\includegraphics[width=0.48\textwidth]{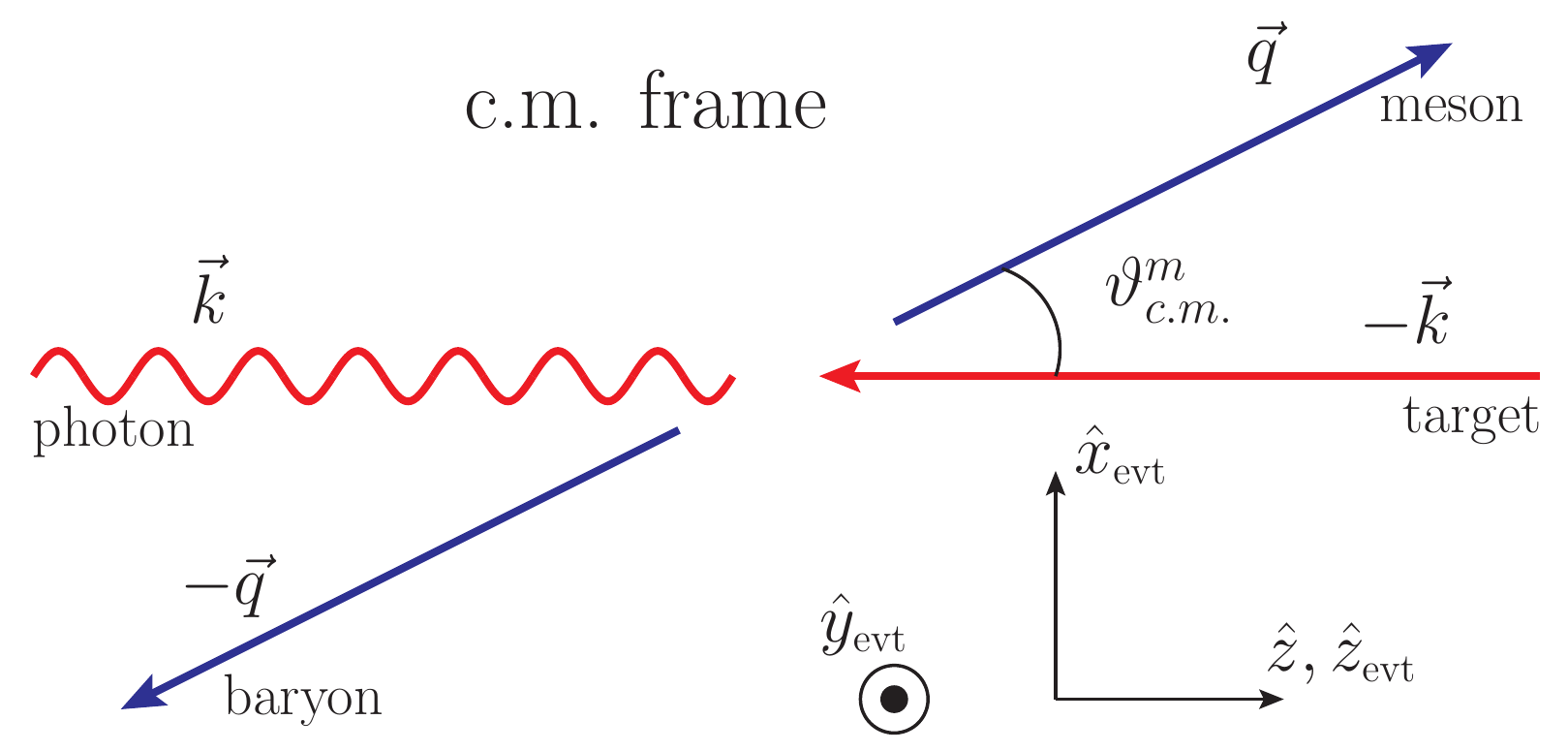}
\caption[]{(Color on-line) Axes for pseudo-scalar meson photoproduction in the center-of-mass (c.m.) frame for a particular event. The $z$-axis is along the photon momentum direction and the $y$-axis is normal to the reaction plane. Momenta of the incoming (outgoing) particles are shown in bold red (blue) arrows. $\vartheta^{m}_{\mbox{\scriptsize c.m.}}$ is the polar meson-production angle in the c.m. frame.  See text for details.\label{fig:axis_setup}} 
\end{figure}

In the case of single pseudo-scalar meson photoproduction, let $\vec{k}$, $-\vec{k}$, $\vec{q}$ and $-\vec{q}$ be the momenta of the incoming photon, target baryon, outgoing meson, and outgoing baryon, respectively, in the overall center-of-mass (c.m.) frame (see Fig.~\ref{fig:axis_setup}). The beam direction defines the $z$-axis, $\hat{z}_{\mbox{\scriptsize evt}} = \vec{k}/|\vec{k}|$. The $y$-axis is taken to be normal to the reaction plane established by the photon and meson momenta, $\hat{y}_{\mbox{\scriptsize evt}} = \vec{k} \times \vec{q}/ |\vec{k} \times \vec{q}|$. 
The $x$-axis is then simply $\hat{x}_{\mbox{\scriptsize evt}} = \hat{y}_{\mbox{\scriptsize evt}} \times \hat{z}_{\mbox{\scriptsize evt}}$.  Here the subscript ``evt'' denotes that these axes, with $\hat{x}_{\mbox{\scriptsize evt}}$ and $\hat{y}_{\mbox{\scriptsize evt}}$ parallel and perpendicular to the reaction plane, are defined on an event-by-event basis.

\section{The photon polarization state and density matrix}

There is some disparity between the optics and particle-physics community in the nomenclature of the right- and left-handed polarization states.  Particle physicists define the right-handed polarization state following the right-hand rule for the transverse electric polarization vector. The spin of the photon points along its momentum for the right-handed polarization state (or positive helicity state). The left-handed polarization state has the photon spin anti-parallel to its direction of motion. The optics community swaps the definitions for the right- and left-handed states, though the notions of positive- and negative-helicity states are the same in both treatments. Here, we adhere to the particle-physics convention.

We will define the polarization basis states for the photon as
\begin{subequations}
\begin{eqnarray}\label{eqn:photon_pols}
|\epsilon^{+}_{\mbox{\scriptsize evt}}\rangle &=& -(|\hat{x}_{\mbox{\scriptsize evt}}\rangle + i |\hat{y}_{\mbox{\scriptsize evt}}\rangle) /\sqrt{2} \\
|\epsilon^{-}_{\mbox{\scriptsize evt}}\rangle &=& (|\hat{x}_{\mbox{\scriptsize evt}}\rangle - i |\hat{y}_{\mbox{\scriptsize evt}}\rangle) /\sqrt{2},
\end{eqnarray}
\end{subequations}
where $|\epsilon^{+}\rangle$ is the right-handed (positive-helicity) state, $|\epsilon^{-}\rangle$ is the left-handed (negative-helicity) state, and $|\hat{x}_{\mbox{\scriptsize evt}}\rangle$ and $|\hat{y}_{\mbox{\scriptsize evt}}\rangle$ are states of transverse polarization along $\hat{x}_{\mbox{\scriptsize evt}}$ and $\hat{y}_{\mbox{\scriptsize evt}}$, respectively. Looking {\em into} the incoming beam, the $y$-component phase leads (trails) the $x$-component phase for the positive (negative) helicity states and the polarization vector rotates counter-clockwise (clockwise) for the positive (negative) helicity states, in accordance with the right-hand rule. 

For a general mixed state, it is useful to switch to the density matrix notation for describing the polarization state of the photon. We follow the work of Adelseck and Saghai (AS)~\cite{as} and FTS~\cite{fts} and write the photon spin density matrix as
\begin{eqnarray}
\rho^\gamma \!\!&=& \!\!\frac{1}{2} \left[\begin{array}{cc} 1 + P^\gamma_C & - P^\gamma_L \exp(-2i(\theta - \varphi)) \\ - P^\gamma_L \exp(2i(\theta - \varphi)) & 1 - P^\gamma_C \end{array}\right]_{\scriptsize \mbox{AS}} \nonumber \\
\!\!&=&\!\! \frac{1}{2}  \left[\begin{array}{cc} 1 + P^S_z & P^S_x -i P^S_y  \\  P^S_x +i P^S_y & 1 - P^S_z \end{array}\right]_{\scriptsize \mbox{FTS}}.
\end{eqnarray}
In the AS prescription, the quantities $P^\gamma_L$ and $P^\gamma_C$ denote the degree of linear ($L$) and circular ($C$) polarization. 
In the FTS treatment, $\vec{P}^{S}$ is the Stokes' vector common in optics, with $x$-, $y$-, and $z$-components indicating the amount of polarization along each spatial direction. 
The kinematic variable $\varphi$ is the azimuthal angle between the reaction plane and the laboratory $\hat{x}_{\mbox{\scriptsize lab}}$ for a linearly polarized photon for a given event. 
The linear polarization vector relative to the laboratory frame is
\begin{equation}
\hat{n}_{\mbox{\scriptsize lab}} = \cos \theta\; \hat{x}_{\mbox{\scriptsize lab}} + \sin \theta\; \hat{y}_{\mbox{\scriptsize lab}}.
\end{equation}

\begin{figure}
\centering
\includegraphics[width=0.5\textwidth]{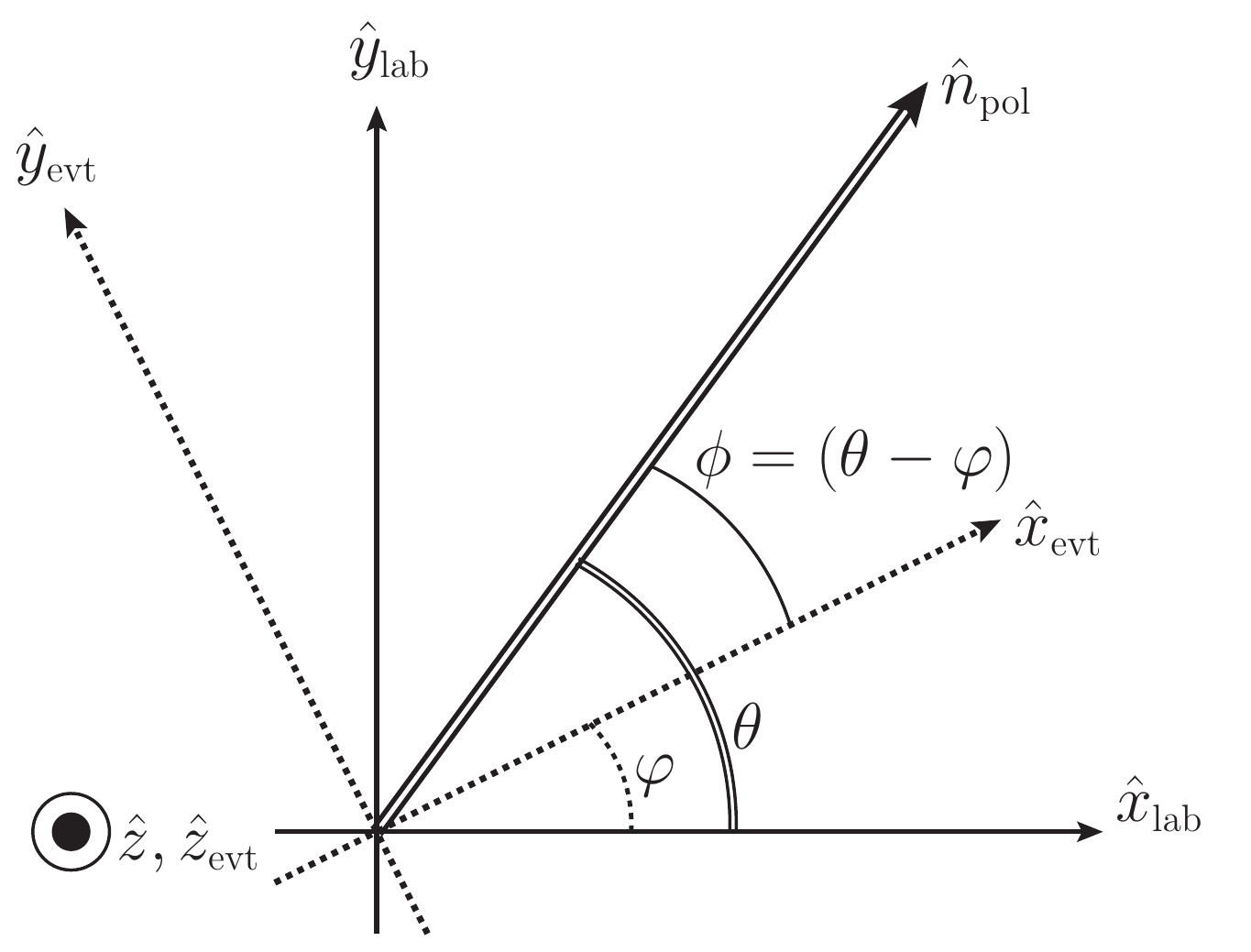}
\caption[]{For a linearly polarized beam, $\{ \hat{x}_{\mbox{\scriptsize lab}}, \hat{y}_{\mbox{\scriptsize lab}}, \hat{z}_{\mbox{\scriptsize lab}} \}$ defines the fixed laboratory axis of the experiment. The polarization direction ($\hat{n}_{\scriptsize \mbox{pol}}$) is at an angle $\theta$ to the laboratory $x$-axis. For a given event, the reaction plane is at angle $\varphi$ to the laboratory $x$-axis. Therefore, the polarization vector is at an angle $\phi = (\theta - \varphi)$ relative to the reaction plane.\label{fig:pol_angles}
} 
\end{figure}

For the circularly polarized photon (or unpolarized beam) case, the production amplitude is azimuthally symmetric about the beam direction. However, for the linearly polarized case, $\hat{x}_{\mbox{\scriptsize lab}}$ and $\hat{y}_{\mbox{\scriptsize lab}}$ define a preferred transverse coordinate system. The experimental conditions define $\hat{n}_{\mbox{\scriptsize lab}}$ and $\theta = 0^\circ$ ($90^\circ$) correspond to parallel (perpendicular) plane polarizations according to the experimentalist. For a given event, relative to the reaction plane, the polarization vector is
\begin{equation}
\hat{n}_{\mbox{\scriptsize evt}} = \cos (\theta - \varphi)\hat{x}_{\mbox{\scriptsize evt}} + \sin (\theta - \varphi) \hat{y}_{\mbox{\scriptsize evt}},
\end{equation}  
and the Stokes' vector $\vec{P}^S$ may be related to the circular polarization quantities by 
\begin{subequations}
\begin{eqnarray}
P^S_z &=& P^\gamma_C \\
P^S_x &=& -P^\gamma_L \cos (2 \phi) \\
P^S_y &=& -P^\gamma_L \sin (2\phi)
\end{eqnarray}
\end{subequations}
in the basis formed by $\{ \hat{x}_{\mbox{\scriptsize evt}}, \hat{y}_{\mbox{\scriptsize evt}}, \hat{z}_{\mbox{\scriptsize evt}} \}$ and $\phi = (\theta - \varphi)$. The angles $\varphi$, $\theta$, and $\phi$ are shown schematically in Fig.~\ref{fig:pol_angles} and the connection between $\vec{P}^S$ and the different polarization states are given in Table~\ref{table:pol_stokes}. Apart from the right- ($r$) and left-handed ($l$) circular polarizations, which are our basis states, there are the perpendicular ($\perp$) and parallel ($\parallel$) states, corresponding to photons linearly polarized along the $y$- and $x$-axes, respectively, and two linearly polarized states at $\pm 45^\circ$ to the $x$-axis (in the $\hat{x}$-$\hat{y}$ plane), labelled as $\pm t$.

\begin{table}[t]
  \begin{center}
  \begin{tabular}{|l|c|c|c|} \hline \hline
   Polarization                                      & $P^S_x$  &   $P^S_y$   &   $P^S_z$  \\ \hline \hline
   ($r$) Circular helicity +1                    & 0        &     0       &    +1      \\
   ($l$) Circular helicity -1                       & 0        &     0       &    -1      \\ \hline
   ($\perp$) Linear ($\phi = \pi/2$)     & +1       &     0       &    0       \\
   ($\parallel$) Linear ($\phi = 0$)      & -1       &     0       &    0       \\ \hline
   ($-t$) Linear ($\phi = -\pi/4$)     & 0       &     +1       &    0       \\
   ($+t$) Linear ($\phi = \pi/4$)      & 0       &     -1       &    0       \\ \hline
   \end{tabular}
   \caption{Stokes' vector $\vec{P}^S$ for different photon polarization configurations (adapted from Ref.~\cite{fts}). The right- ($r$) and left-handed ($l$) circular polarizations are our basis states. The different configurations for the linearly polarized states can be expressed in terms of these basis states. $\phi$ is the angle the linear polarization direction makes with $\hat{x}_{\mbox{\scriptsize evt}}$. See text for details.}
   \label{table:pol_stokes}
 \end{center}
\end{table}

\section{The intensity profile and $T_{lmn}$ elements}

We first note that as far as the \textit{theoretical} definitions of the polarization observables are concerned, the only relevant kinematic variables for a given total energy ($W$) in the c.m.~frame are the angle between the photon polarization vector and the reaction plane, $\phi = (\theta - \varphi)$ (for a linearly polarized beam), and the polar meson production angle $\vartheta^{m}_{\mbox{\scriptsize c.m.}}$. 
This is because the observables are defined as asymmetries relative to the reaction-plane coordinate system $\{\hat{x}_{\mbox{\scriptsize evt}}, \hat{y}_{\mbox{\scriptsize evt}}, \hat{z}_{\mbox{\scriptsize evt}} \}$. It is only when we will need to connect the observables to experimentally measurable intensity distributions, that the orientation between reaction plane, photon polarization vector, and the lab frame (quantified by angles $\theta$ and $\varphi$) will be required. 
We will return to this point later in Sec.~\ref{sec:experimental_profiles}. In what follows (as in the FTS conventions) we will refer to the Pauli matrices operating in the spin spaces of the beam, target baryon, and outgoing baryon as $\sigma^\gamma$, $\sigma^i$, and $\sigma^b$, respectively. The density matrices are then given by 
\begin{subequations}
\begin{eqnarray}
\rho^\gamma &=& \frac{1}{2}( 1 + \vec{P}^S \cdot \vec{\sigma}^\gamma) \\
\rho^i &=& \frac{1}{2}( 1 + \vec{P}^i \cdot \vec{\sigma}^i) \\
\rho^b &=& \frac{1}{2}( 1 + \vec{P}^b \cdot \vec{\sigma}^b).
\end{eqnarray}
\label{eqn:den_matrices}
\end{subequations}
The vectors $\vec{P}^S$, $\vec{P}^i$, and $\vec{P}^b$ denote the polarization vectors of the beam, target and recoiling baryon, respectively. If any of the beam, target or recoil polarization is not measured, the corresponding $\vec{P}$ is the zero vector. 

We define $\mathcal{A}_{m_\gamma m_i m_b}$ to be the reaction amplitude for a particular spin configuration of the photon ($m_\gamma$), target ($m_i$) and baryon ($m_b$), with the spin-quantization axes for all particles along $\hat{z}$, the beam direction.
For a given photon spin, $m_{\gamma}$, the four $\mathcal{A}$ amplitudes correspond to the elements of a $2 \times 2$ matrix $J_{m_\gamma}$ in the space of transition elements for a target spin state $|m_i\rangle$ going to the baryon spin state $|m_b \rangle$. 
Therefore the matrix elements of $J_{m_\gamma}$ are $(J_{m_\gamma})_{m_b m_i} = \langle m_b| J_{m_\gamma}| m_i \rangle$ and $J$ and $\mathcal{A}$ are connected via 
\begin{equation}
\langle m_b| J_{m_\gamma}| m_i \rangle = \mathcal{A}_{m_\gamma m_i m_b}.
\end{equation}

For experiments with ``mixed'' states $\rho_{\mbox{\scriptsize in}}$ and $\rho_{\mbox{\scriptsize out}}$ as the initial prepared (input) and final measured (output) configurations connected by the transition operator $J$, the intensity profile is proportional to the trace $\mbox{Tr}[{\rho_{\mbox{\scriptsize out}} J \rho_{\mbox{\scriptsize in}} J^\dagger}]$. In the present case, $\rho_{\mbox{\scriptsize in}} = \rho^i \otimes \rho^\gamma$ and $\rho_{\mbox{\scriptsize out}} = \rho^b$. Therefore, the most general intensity expression for the profile dependent upon beam, target, and recoil polarizations is given by
\begin{equation}
\sigma  = \sigma_0 \left(\frac{\mbox{Tr}[\rho^b J \rho^i \rho^\gamma J^\dagger]}{\mbox{Tr}[J J^\dagger]}\right),
\label{eqn:full_profile}
\end{equation}
where $\sigma_0$ is the unpolarized cross-section and the traces are over the beam, target and recoil spins. This derivation can be found in FTS~\cite{fts} and in an equivalent form, in the paper by Goldstein, {\em et al}~\cite{goldstein}. The main utility of this formulation is its symbolic compactness which  enables easy derivation of other observables and their correlations. In Sec.~\ref{sec:computation} we will describe in detail how to expand these traces to give amplitude-level expressions for the polarization observables.\\

We now establish a notation for the Pauli matrices and the polarization vectors as four-component vectors, wherein
\begin{subequations}
\begin{eqnarray}
\{\sigma_0, \sigma_1, \sigma_2, \sigma_3\} &\equiv& \{I,\sigma_x, \sigma_y, \sigma_z\} \\
\{P_0, P_1, P_2, P_3\} &\equiv& \{1, P_x, P_y, P_z\}. 
\end{eqnarray}
\end{subequations} 
Since each density matrix in Eqs.~\ref{eqn:den_matrices} has four terms, the full profile in Eq.~\ref{eqn:full_profile} has $4\times4\times 4 = 64$ terms. We will also adopt the convention:
\begin{equation}
T_{lmn} \equiv \frac{\mbox{Tr}[\sigma^b_{n} J \sigma^i_m \sigma^\gamma_l J^\dagger]}{\mbox{Tr}[J J^\dagger]},
\end{equation} 
so that Eq.~\ref{eqn:full_profile} can be compactly represented as 
\begin{equation}
\mathcal{I} \;\;\sim \displaystyle \sum_{l m n \;\in\; \{0,1,2,3\} } P^S_l P^i_m P^b_n\;  T_{lmn}.
\label{eqn:profile_lmn}
\end{equation}

\section{Mirror symmetry transformations}
\label{sec:mirror_symmetry}
Following the work of Artru {\em et al}~\cite{artru}, we first define the mirror inversion operator
\begin{equation}
M = \Pi \; \mbox{exp}(-i\pi J_y),
\end{equation}
which describes a parity inversion $(\Pi)$ followed by a $180^\circ$ rotation about the $\hat{y}$ axis. We list the effects of $M$ on the relevant particle types labeled by their spin and parity quantum numbers, $J^{P}$:
\begin{itemize}
\item{$J^{P} = 0^{-}$ pseudo-scalar meson:} Only the parity inversion contributes.  Thus, $M = \Pi = -1$ is a simple sign flip.
\item{$J^{P} = \frac{1}{2}^{+}$ baryons:} Here $\Pi = 1$ and $J_y = \sigma_2/2$. Therefore ${M = \Pi \; \mbox{exp}(-i\pi J_y) = -i\sigma_2}$. In terms of the spin states, the transformation is given by $|+\rangle \to |-\rangle$ and $|-\rangle \to - |+\rangle$.
\item{$J^{P} = 1^{-}$ photon:} A $180^\circ$ rotation about the $\hat{y}$ axis leaves $|\hat{y}\rangle$ unchanged but changes $|\hat{x}\rangle$ to $|-\hat{x}\rangle$. Substituting this in Eq.~\ref{eqn:photon_pols} leads to an interchange between $|\epsilon^{+}_{\mbox{\scriptsize evt}}\rangle$ and $|\epsilon^{-}_{\mbox{\scriptsize evt}}\rangle$. Including $\Pi = -1$ for a vector particle leads to $M = -\sigma_1$ for the photon in the Pauli basis. In terms of the spin states, the transformation is $|\epsilon^{+}_{\mbox{\scriptsize evt}}\rangle \leftrightarrow - |\epsilon^{-}_{\mbox{\scriptsize evt}}\rangle$. 
\end{itemize}

There are two main effects of the $M$ transformation of which we make use. First, $M$ acting on any $T_{lmn}$ element results in a reshuffle in the Pauli operators for the incoming states:
\begin{eqnarray}
\{0,1,2,3\} &\Rightarrow& \{-1,\,-0,-i3,\;i2 \} \hspace{0.2cm} \mbox{(photon)}\\
\label{eqn:pi_reshuffle_photon}
\{0,1,2,3\} &\Rightarrow& \{-i2,-3,-i0,\;1 \} \hspace{0.2cm} \mbox{(target)}
\label{eqn:pi_reshuffle_baryon}.
\end{eqnarray}
However, Eq.~\ref{eqn:pi_reshuffle_baryon} does not quite work for the outgoing baryon density matrix, since the effect of the pseudo-scalar meson in the outgoing system needs to be incorporated as well. For the outgoing meson-baryon system, the reshuffle is given by 
\begin{equation}
\{0,1,2,3\} \Rightarrow \{-i2,3,-i0,\;-1 \} \hspace{0.2cm} \mbox{(outgoing system)}
\label{eqn:pi_reshuffle_outgoingbaryon},
\end{equation}
where we have added an extra sign flip for $\sigma_1$ and $\sigma_3$ compared to Eq.~\ref{eqn:pi_reshuffle_baryon}, that comes from the parity of the pseudo-scalar meson. The $\sigma_0$ terms do not acquire this extra sign flip, since they physically correspond to the situation where the experiment is ``blind'' to the spins of the outgoing states. Also, since $\sigma_2$ is connected to the identity matrix by the $M$ transform, it does not acquire a sign flip. 

Second, the action of $M$ on the production amplitudes and invariance under this transformation lead to relations between the amplitudes for positive and negative photon helicities:
\begin{subequations}
\begin{eqnarray}
L_1 &\equiv& \mathcal{A}_{+++} = +\mathcal{A}_{---} \\
L_2 &\equiv& \mathcal{A}_{++-} = -\mathcal{A}_{--+} \\
L_3 &\equiv& \mathcal{A}_{+--} = +\mathcal{A}_{-++} \\
L_4 &\equiv& \mathcal{A}_{+-+} = - \mathcal{A}_{-+-},
\end{eqnarray}
\end{subequations}
where the four independent amplitudes $L_i$ will be called the {\em longitudinal basis amplitudes}. \\

The $L_i$ amplitudes are very closely related to the standard CGLN amplitudes~\cite{cgln}, since they are both in the Cartesian basis. In the CGLN approach, one writes the amplitude (up to over all phase and energy factors) as 
\begin{equation}
\mathcal{A}_{\lambda m_i m_b} \sim \chi^\dagger(m_b)\mathcal{F}(\lambda) \chi(m_i),
\end{equation} 
where $\chi(m_i)$ and $\chi(m_b)$ are the spinors of the initial target and the final baryon, respectively, and $\lambda = \pm 1$ gives the photon helicity. The matrix $\mathcal{F}(\lambda)$ is expanded in terms of the four CGLN amplitudes $F_i$ as
\begin{eqnarray}
\mathcal{F}(\lambda) &=& i(\vec{\sigma} \cdot \hat{\epsilon}) F_1 + (\hat{\sigma}\cdot \hat{q})(\hat{\sigma} \times \hat{k})\cdot \hat{\epsilon} F_2 \nonumber \\
& & +\; i (\hat{\epsilon} \cdot \hat{q}) (\vec{\sigma}\cdot \hat{k}) F_3 + i (\hat{\epsilon} \cdot \hat{q} )(\vec{\sigma}\cdot \hat{q}) F_4.
\end{eqnarray}
For $\lambda =+1$, this leads to 
\begin{equation}
\mathcal{F}(+1) \sim  \left[\begin{array}{cc} s F_3 + cs F_4 & 2F_1 - 2cF_2 + s^2F_4 \\ s^2 F_4 & -2sF_2 -sF_3 -csF_4\end{array}\right],
\end{equation}
where we have abbreviated ${s = \sin(\vartheta^m_{\mbox{\scriptsize c.m.}})}$ and ${c = \cos(\vartheta^m_{\mbox{\scriptsize c.m.}})}$ in terms of the polar meson production angle in the c.m.~frame, $\vartheta^m_{\mbox{\scriptsize c.m.}}$ (see Fig.~\ref{fig:axis_setup}). The connections between the $L_i$ amplitudes and the CGLN amplitudes $F_i$ are therefore:
\begin{subequations}
\begin{eqnarray}
L_1 &\sim& sF_3 + cs F_4 \\
L_2 &\sim& s^2 F_4 \\
L_3 &\sim& -2sF_2 -sF_3 -csF_4\\
L_4 &\sim& 2F_1 - 2cF_2+s^2F_4.
\end{eqnarray}
\end{subequations}

\begin{table}[t]
  \begin{center}
   \begin{tabular}{|c|c|c|c|} \hline \hline
    Type & Observable & Definition & $M$ transform \\ \hline \hline
    Unpolarized & $\sigma_0$ & $(000)$ & $(122)$\\ \hline \hline
    Single-pol. & $P$ & $(002)$ & $(120)$\\ \hline
    '' & $\Sigma$ & $(100)$ & $(022)$\\ \hline
    '' & $T$ & $(020)$ & $(102)$\\ \hline \hline 
    Beam-target & $E$ & $(330)$ & $(212)$\\ \hline
    '' & $F$ & $(310)$ & -$(232)$\\ \hline
    '' & $G$ & -$(230)$ & $(312)$\\ \hline
    '' & $H$ & -$(210)$ & -$(332)$\\ \hline \hline
    Beam-recoil &$C_x$ & $(301)$ & $(223)$\\ \hline
    '' & $C_z$ & $(303)$ & -$(221)$\\ \hline
    '' & $O_x$ & -$(201)$ & $(323)$\\ \hline
    '' & $O_z$ & -$(203)$ & -$(321)$\\ \hline \hline
    Target-recoil & $T_x$ & $(011)$ & $(133)$\\ \hline
    '' & $T_z$ & $(013)$ & -$(131)$\\ \hline
    '' & $L_x$ & $(031)$ & -$(113)$\\ \hline
    '' & $L_z$ & $(033)$ & $(111)$\\ \hline \hline
    \end{tabular}
  \end{center}
\caption[]{The definition of the 16 observables as the $T_{lmn}$ correlations in Eq.~\ref{eqn:profile_lmn}. The defining $T_{lmn}$ elements are listed as $(lmn)$ in the third column and the corresponding $M$ transformed elements are listed in the last column. Invariance under the $M$ transform results in each observable occurring twice. The full intensity expansion is given in Eq.~\ref{eqn:full_profile_expanded}.}
\label{table:obs}
\end{table}

\section{The Polarization observables}

We first define the sixteen observables as the various $T_{lmn}$ elements (see Table~\ref{table:obs}) noting that each observable occurs twice in the expansion given by Eq.~\ref{eqn:profile_lmn}. Our definitions for the observables follow those in FTS~\cite{fts}. It is to be noted that there are minus signs in front of the defining $T_{lmn}$ elements for the four double-polarization observables that use a linearly polarized beam ($G$, $H$, $O_x$, and $O_z$). These extra sign flips are needed to preserve the definitions of these variables as {\em physical asymmetries}, as given in FTS~\cite{fts}. The signs for $G$, $H$, $O_x$, and $O_z$, in terms of the {\em density matrix trace calculations}, as given in Appendix A in the FTS article will therefore acquire sign flips (see Sec.~\ref{sec:beam_target_profile} and Sec.~\ref{sec:beam_recoil_profile} for further details). We now write the full intensity profile in terms of the polarization observables as:
\begin{widetext}
\begin{eqnarray}
\mathcal{I} &=& \mathcal{I}_0 \;\{(1+P^S_xP^i_yP^b_y) + P(P^b_y + P^S_xP^i_y) + \Sigma(P^S_x + P^i_y P^b_y) + T(P^i_y + P^S_x P^b_y) + E(P^S_zP^i_z + P^S_y P^i_x P^b_y) \nonumber \\
&& + F(P^S_zP^i_x - P^S_y P^i_zP^b_y) + G(-P^S_yP^i_z + P^S_zP^i_xP^b_y) + H (-P^S_yP^i_x - P^S_z P^i_z P^b_y) +C_x(P^S_zP^b_x + P^S_y P^i_y P^b_z) \nonumber \\
&& + C_z (P^S_zP^b_z - P^S_y P^i_y P^b_x) + O_x (-P^S_y P^b_x + P^S_zP^i_yP^b_z) + O_z( -P^S_yP^b_z -P^S_z P^i_y P^b_x) +T_x(P^i_x P^b_x +P^S_x P^i_z P^b_z) \nonumber \\
&&  + T_z (P^i_x P^b_z - P^S_x P^i_z P^b_x) + L_x(P^i_z P^b_x - P^S_xP^i_xP^b_z) + L_z(P^i_z P^b_z + P^S_x P^i_x P^b_x)\}.
\label{eqn:full_profile_expanded}
\end{eqnarray}
\end{widetext}
We note that the three single polarizations ($P$, $\Sigma$ and $T$) occur again as double correlations and the twelve double polarizations ($E$, $F$, $G$, $H$, $C_x$, $C_z$, $O_x$, $O_z$, $T_x$, $T_z$, $L_x$ and $L_z$) occur again as triple correlations.

\subsection{The 32 vanishing terms}
The expansion in Eq.~\ref{eqn:profile_lmn} has 64 terms, while Eq.~\ref{eqn:full_profile_expanded} has only 32 terms. The rest of the 32 terms vanish under $M$ invariance. $T_{001}$ and $T_{003}$ are examples of such terms (they do not occur in Table~\ref{table:obs}). Physically, these two elements correspond to recoil polarizations (with unpolarized beam and target) along the $\hat{x}$ and $\hat{z}$ directions, which are required by $M$ invariance to be zero. 
The general structure of these vanishing terms can be understood from the following example. From Eq.~\ref{eqn:pi_reshuffle_photon} for the photon, under a $M$ transform, $\sigma_1$ is connected to the the identity matrix ($\sigma_0$). Similarly, from Eqs.~\ref{eqn:pi_reshuffle_baryon} and~\ref{eqn:pi_reshuffle_outgoingbaryon} for the baryons, it is $\sigma_2$ that is connected to the identity matrix. We group $\sigma_0$ and the Pauli matrix connected to $\sigma_0$ by the $M$ operator as ``E'' (type +1), and the rest ($\sigma_2$ and $\sigma_3$ for the photon, and $\sigma_1$ and $\sigma_3$ for the baryons) as of the ``O'' (type -1). A general correlation $T_{lmn}$ vanishes if the product of the ``types'' of $l$, $m$, and $n$ is -1, since these are not invariant under the mirror symmetry transformation.

\subsection{Beam-target type experiments}
\label{sec:beam_target_profile}

We will show that our expressions for the intensity profiles as measured by the experimentalist conform to the definition of these observables as asymmetries. Following the notation set up in FTS~\cite{fts} we will denote the cross-section for any configuration of the beam, target and recoil polarizations as $\sigma^{(\gamma,i,b)}$. For beam-target type experiments, $\vec{P}^{b} = \vec{0}$, and Eq.~\ref{eqn:full_profile_expanded} becomes
\begin{eqnarray}
\sigma^{(\gamma, i, 0)} &=& \sigma_0 \{1 + P^S_x \Sigma + P^i_x(-P^S_y H + P^S_z F) \nonumber \\
&& + P^i_y(T + P^S_x P) + P^i_z(-P^S_y G + P^S_z E) \}.
\end{eqnarray}
The beam asymmetry is defined as 
\begin{equation}
\Sigma  = \frac{\sigma^{(\perp,0,0)} - \sigma^{(\parallel,0,0)}}{\sigma^{(\perp,0,0)} + \sigma^{(\parallel,0,0)}},
\label{eqn:beam_asym}
\end{equation}
where $\parallel$ and $\perp$ correspond to a beams with polarizations along the $\hat{x}_{\mbox{\scriptsize evt}}$ ($\phi = 0$) and $\hat{y}_{\mbox{\scriptsize evt}}$ ($\phi = \pi/2$) directions, respectively, and $0$  denotes an unpolarized configuration. The target asymmetry is defined as 
\begin{equation}
T  = \frac{\sigma^{(0,+y,0)} - \sigma^{(0,-y,0)}}{\sigma^{(0,+y,0)} + \sigma^{(0,-y,0)}},
\label{eqn:targ_asym}
\end{equation}
and the four double polarizations are defined as
\begin{subequations}
\label{eqn:beam_target_defs}
\begin{eqnarray}
E &=& \frac{\sigma^{(r,+z,0)} - \sigma^{(r,-z,0)}}{\sigma^{(r,+z,0)} + \sigma^{(r,-z,0)}} \\
F &=& \frac{\sigma^{(r,+x,0)} - \sigma^{(r,-x,0)}}{\sigma^{(r,+x,0)} + \sigma^{(r,-x,0)}} \\
G &=& \frac{\sigma^{(+t,+z,0)} - \sigma^{(+t,-z,0)}}{\sigma^{(+t,+z,0)} + \sigma^{(+t,-z,0)}} \\
H &=& \frac{\sigma^{(+t,+x,0)} - \sigma^{(+t,-x,0)}}{\sigma^{(+t,+x,0)} + \sigma^{(+t,-x,0)}},
\end{eqnarray}
\end{subequations} 
where ``$r$'' denotes a right-handed circularly polarized beam (all photons in the state $|\epsilon^{+}_{\mbox{\scriptsize evt}}\rangle$), and ``$+t$'' denotes a linearly polarized beam with $\phi = +\pi/4$ with respect to $\hat{x}_{\mbox{\scriptsize evt}}$. The full expression for the cross-section in beam-target experiments reads
\begin{eqnarray}
\sigma^{(\gamma, i, 0)}_{\mbox{\scriptsize theory}} &=& \sigma_0 \{ 1 - P^\gamma_L \Sigma \cos(2 \phi) + P^i_y \left(T-P^\gamma_L P \sin (2 \phi)\right) \nonumber  \\
  & & \hspace{1.5cm} + P^i_x \left(P^\gamma_C F + P^\gamma_L H \sin (2 \phi)\right) \nonumber \\
&& \hspace{1.6cm} + P^i_z \left( P^\gamma_C E + P^\gamma_L G \sin (2 \phi) \right) \}
\label{eqn:beam_target_profile}
\end{eqnarray}
where we have added a subscript ``theory'' to remind the reader that this is for the theoretical formalism only. It can easily be checked that the definitions in Eqs.~\ref{eqn:beam_asym}-\ref{eqn:beam_target_defs} are consistent with the intensity profile given by Eq.~\ref{eqn:beam_target_profile}. Recall that $\phi = +\pi/4$ corresponds to $P^S_y = -1$ (see Table~\ref{table:pol_stokes}), explaining the extra minus signs for $G$ and $H$ in the definitions of the corresponding $T_{lmn}$ elements in Table~\ref{table:obs}. 

It is to be noted that one has access to an ``extra'' single-polarization observable, the recoil polarization $P$, even though the polarization of the recoiling baryon is not measured here. This is again due to the $M$ transform relations. In fact, any double-polarization experiment has access to all the three single polarization observables. The definition of $P$ as an asymmetry is given in the next sub-section.

\subsection{Beam-recoil type experiments}
\label{sec:beam_recoil_profile}

For an experiment with beam and recoil baryon polarization information, we follow a similar logic.  Here, $\vec{P}^{i} = \vec{0}$, and the beam-recoil expression is 
\begin{eqnarray}
\sigma^{(\gamma, 0, b)}_{\mbox{\scriptsize theory}} &=& \sigma_0 \{ 1 - P^\gamma_L \Sigma \cos(2 \phi) + P^b_y (P - P^\gamma_L T \cos(2\phi) ) \nonumber \\ 
 & &+ P^b_x \left(P^\gamma_C C_x + P^\gamma_L O_x \sin (2 \phi)\right) \nonumber \\
& & + P^b_z \left( P^\gamma_C C_z + P^\gamma_L O_z \sin (2 \phi) \right) \},
\label{eqn:beam_recoil_profile}
\end{eqnarray}
where the recoil polarization $P$ is defined as
\begin{equation}
P  = \frac{\sigma^{(0,0,+y)} - \sigma^{(0,0,-y)}}{\sigma^{(0,0,+y)} + \sigma^{(0,0,-y)}}.
\end{equation}
The four beam-recoil double polarizations are 
\begin{subequations}
\label{eqn:beam_recoil_obs}
\begin{eqnarray}
C_z &=& \frac{\sigma^{(r,0,+z)} - \sigma^{(r,0,-z)}}{\sigma^{(r,0,+z)} + \sigma^{(r,0,-z)}} \\
C_x &=& \frac{\sigma^{(r,0,+x)} - \sigma^{(r,0,-x)}}{\sigma^{(r,0,+x)} + \sigma^{(r,0,-x)}} \\
O_z &=& \frac{\sigma^{(+t,0,+z)} - \sigma^{(+t,0,-z)}}{\sigma^{(+t,0,+z)} + \sigma^{(+t,0,-z)}} \\
O_x &=& \frac{\sigma^{(+t,0,+x)} - \sigma^{(+t,0,-x)}}{\sigma^{(+t,0,+x)} + \sigma^{(+t,0,-x)}}.
\end{eqnarray}
\end{subequations} 
The ``extra'' single polarization observable accessible here is the target asymmetry $T$, defined in Eq.~\ref{eqn:targ_asym}.

As in the case of $G$ and $H$, the definitions of $O_x$ and $O_z$ as asymmetries use $\phi = +\pi/4$ that corresponds to $P^S_y = -1$. This explains the extra minus signs in the defining $T_{lmn}$ elements in Table~\ref{table:obs} for $O_x$ and $O_z$.

\subsection{Target-recoil type experiments}
\label{sec:target_recoil_profile}

The target-recoil expression is 
\begin{eqnarray}
\sigma^{(0, i, b)} &=& \sigma^0 \{ 1 + P^i_y T + P^b_y (P+ \Sigma P^i_y) \nonumber \\
& & + P^i_z \left( P^b_z L_z + P^b_x L_x  \right) +  P^i_x \left( P^b_x T_x + P^b_z T_z  \right) \} \nonumber \\,
\label{eqn:target_recoil_profile}
\end{eqnarray}
where the four target-recoil double polarizations are 
\begin{subequations}
\label{eqn:target_recoil_obs}
\begin{eqnarray}
T_z &=& \frac{\sigma^{(0,+x,+z)} - \sigma^{(0,+x,-z)}}{\sigma^{(0,+x,+z)} + \sigma^{(0,+x,-z)}} \\
T_x &=& \frac{\sigma^{(0,+x,+x)} - \sigma^{(0,+x,-x)}}{\sigma^{(0,+x,+x)} + \sigma^{(0,+x,-x)}} \\
L_z &=& \frac{\sigma^{(0,+z,+z)} - \sigma^{(0,+z,-z)}}{\sigma^{(0,+z,+z)} + \sigma^{(0,+z,-z)}} \\
L_x &=& \frac{\sigma^{(0,+z,+x)} - \sigma^{(0,+z,-x)}}{\sigma^{(0,+z,+x)} + \sigma^{(0,+z,-x)}}.
\end{eqnarray}
\end{subequations} 

\section{Connection with experimental intensity profiles}
\label{sec:experimental_profiles}
Until now, we have been careful to distinguish the ``theoretical'' intensity profiles from what experimentalists will actually measure. The only difference lies in the case of a linearly polarized beam where the laboratory analyzing direction set by the choice of the angle $\theta$ (see Fig.~\ref{fig:pol_angles}) can vary. 
Eq.~\ref{eqn:target_recoil_profile} remains the same between the theory and experimental formalisms, since it is independent of $\theta$. For Eqs.~\ref{eqn:beam_target_profile} and~\ref{eqn:beam_recoil_profile}, however, we need to get back to the relation $\phi = (\theta - \varphi)$ in Fig.~\ref{fig:pol_angles}. The easiest choice is to measure everything with respect to $\hat{x}_{\mbox{\scriptsize lab}}$, which usually represents the experimentalist's choice of photon polarization axis. Therefore, we take $\theta = 0$ (also called ``para'' setting), so that $\phi = - \varphi$ and Eqs.~\ref{eqn:beam_target_profile} and~\ref{eqn:beam_recoil_profile} become
\begin{eqnarray}
\sigma^{(\gamma, i, 0)}_{\mbox{\scriptsize para}} &=& \sigma_0 \{ 1 - P^\gamma_L \Sigma \cos(2 \varphi) + P^i_y \left(T+P^\gamma_L P \sin (2 \varphi)\right) \nonumber  \\
  & & + P^i_x \left(P^\gamma_C F - P^\gamma_L H \sin (2 \varphi)\right) \nonumber \\
& & + P^i_z \left( P^\gamma_C E - P^\gamma_L G \sin (2 \varphi) \right) \},
\end{eqnarray}
and 
\begin{eqnarray}
\sigma^{(\gamma, 0, b)}_{\mbox{\scriptsize para}} &=& \sigma_0 \{ 1 - P^\gamma_L \Sigma \cos(2 \varphi) + P^b_y (P - P^\gamma_L T \cos(2\varphi) ) \nonumber \\ 
 & &+ P^b_x \left(P^\gamma_C C_x - P^\gamma_L O_x \sin (2 \varphi)\right) \nonumber \\
& & + P^b_z \left( P^\gamma_C C_z - P^\gamma_L O_z \sin (2 \varphi) \right) \},
\label{eqn:beam_recoil_profile_1}
\end{eqnarray}
respectively. Similarly, for $\theta = \pi/2$ (also called ``perp'' setting), we get
\begin{eqnarray}
\sigma^{(\gamma, i, 0)}_{\mbox{\scriptsize perp}} &=& \sigma_0 \{ 1 + P^\gamma_L \Sigma \cos(2 \varphi) + P^i_y \left(T-P^\gamma_L P \sin (2 \varphi)\right) \nonumber  \\
  & & + P^i_x \left(P^\gamma_C F + P^\gamma_L H \sin (2 \varphi)\right) \nonumber \\
& & + P^i_z \left( P^\gamma_C E + P^\gamma_L G \sin (2 \varphi) \right) \},
\end{eqnarray}
and 
\begin{eqnarray}
\sigma^{(\gamma, 0, b)}_{\mbox{\scriptsize perp}} &=& \sigma_0 \{ 1 + P^\gamma_L \Sigma \cos(2 \varphi) + P^b_y (P + P^\gamma_L T \cos(2\varphi) ) \nonumber \\ 
 & &+ P^b_x \left(P^\gamma_C C_x + P^\gamma_L O_x \sin (2 \varphi)\right) \nonumber \\
& & + P^b_z \left( P^\gamma_C C_z + P^\gamma_L O_z \sin (2 \varphi) \right) \}.
\label{eqn:beam_recoil_profile_2}
\end{eqnarray}
It is important to note that the sine and cosine terms in Eqs.~\ref{eqn:beam_target_profile} and~\ref{eqn:beam_recoil_profile} alter signs differently in going from the ``theory'' expressions to the ``para'' and ``perp'' settings, and this directly affects the signs of the extracted polarization observables. Therefore, care must be taken by the experimentalist to conform to a definition of the ``para'' and ``perp'' settings that matches with the theoretical definitions. Finally, we also note that it is beneficial to measure the intensity profiles for both the ``para'' and ``perp'' settings and extract the polarizations from the asymmetries between the two settings. This removes the overall normalization factor (the unpolarized cross-section), and therefore, any dependence on the detector acceptance (see Ref.~\cite{g8} for details). 

\section{Computation of polarization expressions in the longitudinal basis}
\label{sec:computation}
\subsection{Some basic rules and caveats}

We list some basic caveats that will be useful during the computations.
\begin{enumerate}
\item The matrix representation of an operator is $\mathcal{O}_{nm} = \langle n | \mathcal{O} |m\rangle$ (note order of subscripts). For the Pauli matrices for example, $(\sigma_y)_{+-} = -i$, $(\sigma_z)_{--} = -1$, etc.
\item FTS uses $m_{s'}$ and $m_s$ for outgoing baryon and incoming (target) spins. We adopt $m_b$ and $m_i$ as the final baryon and initial proton spins, respectively, and denote the photon spin by $m_\gamma = \lambda$. Any other index will be a dummy index for summation purposes. Also, unless otherwise mentioned, it is understood that repeated indices are to be summed over. 
\item A useful relation is that for the conjugate operator $J^\dagger_\lambda$, the matrix elements are $(J^\dagger_\lambda)_{m_i m_b} = (J_\lambda)_{m_b m_i}^\ast = \mathcal{A}^\ast_{\lambda m_b m_i}$.
\item There are two types of traces in the FTS paper. ``Tr'' implies a trace over all spins, while ``tr'' implies a trace over the baryon spins, assuming that the photon spins have been traced over. To go from ``Tr'' to ``tr'', that is, the procedure of doing the photon spin trace, is as follows. Let $\Omega_b$ and $\Omega_i$ be any operator in the final baryon and initial target proton spin space respectively. For the three Pauli matrices $\sigma^\gamma_x$, $\sigma^\gamma_y$ and $\sigma^\gamma_z$, the  photon traces are computed as follows:
\begin{subequations}
\begin{eqnarray}
\mbox{Tr}[\Omega_b J \Omega_i \sigma^\gamma_x  J^\dagger] \!&=&\! \displaystyle \sum_{\lambda \lambda'} \mbox{tr}[\Omega_b J_\lambda \Omega_i (\sigma^\gamma_x)_{\lambda \lambda '} (J^\dagger)_{\lambda'}] \nonumber \\
&=& \mbox{tr}[\Omega_b J_+ \Omega_i (J^\dagger)_- + \Omega_b J_- \Omega_i (J^\dagger)_+ ] \nonumber \\ \\
\mbox{Tr}[\Omega_b J \Omega_i \sigma^\gamma_y  J^\dagger] \!&=&\! \displaystyle \sum_{\lambda \lambda'} \mbox{tr}[\Omega_b J_\lambda \Omega_i (\sigma^\gamma_y)_{\lambda \lambda '} (J^\dagger)_{\lambda'}] \nonumber \\
&=& -i\;\; \mbox{tr}[\Omega_b J_+ \Omega_i (J^\dagger)_- - \Omega_b J_- \Omega_i (J^\dagger)_+ ] \nonumber \\ \\
\mbox{Tr}[\Omega_b J \Omega_i \sigma^\gamma_z J^\dagger] \!&=& \!\displaystyle \sum_{\lambda \lambda'} \mbox{tr}[\Omega_b J_\lambda \Omega_i (\sigma^\gamma_z)_{\lambda \lambda '} (J^\dagger)_{\lambda'}] \nonumber \\
&=& \mbox{tr}[\Omega_b J_+ \Omega_i (J^\dagger)_+ - \Omega_b J_- \Omega_i (J^\dagger)_- ]. \nonumber  \\
\end{eqnarray}
\end{subequations}
Note that these expressions for the summations over the photon states are equivalent to those from the more conventional forms that can be found in Ref.~\cite{goldstein}, for example. The trace notation is simply a more compact way of expressing the spin sums.    
\item Overall normalization factor. All 15 polarization observables will be normalized by the intensity factor $\mbox{Tr}[J J^\dagger]$. This is given as
\begin{equation}
\mbox{Tr}[J J^\dagger] = \displaystyle \sum_{\lambda m_i m_b} |\mathcal{A}_{\lambda m_i m_b}|^2.
\end{equation}
This will not appear in our expressions below, but it is understood that this normalization always goes into the computations.
\end{enumerate}

\subsection{The 15 polarization expressions}

The detailed computation of the 15 polarizations are given below: 
\begin{eqnarray}
P \!\!&=\!\!& \mbox{Tr}[\sigma^b_y JJ^\dagger] \nonumber \\
  \!\!&=\!\!& \langle m_b| \sigma_y | m_b'\rangle \langle m_b' |J_\lambda | m_i \rangle \langle m_i |J^\dagger_\lambda | m_b \rangle \nonumber \\
  \!\!&=\!\!& \langle m_b| \sigma_y | m_b'\rangle \langle m_b' |J_\lambda | m_i \rangle (\langle m_b |J_\lambda | m_i \rangle)^\ast \nonumber \\
  \!\!&=\!\!& -i \; \langle - |J_\lambda | m_i \rangle (\langle + |J_\lambda | m_i \rangle)^\ast + i \langle+ |J_\lambda | m_i \rangle (\langle - |J_\lambda | m_i \rangle)^\ast\nonumber \\
  \!\!&=\!\!& \displaystyle \sum_{\lambda m_i}  -2\, \mbox{Im}\left( \mathcal{A}_{\lambda m_i +} \mathcal{A}^\ast_{\lambda m_i -}\right)
\end{eqnarray}

\begin{eqnarray}
\Sigma \!\!&=&\!\! \mbox{Tr}[J \sigma^\gamma_x J^\dagger] \nonumber \\
  \!\!&=&\!\! \langle m_b |J_+ | m_i \rangle \langle m_i |J^\dagger_- | m_b \rangle + \langle m_b |J_- | m_i \rangle \langle m_i |J^\dagger_+ | m_b \rangle \nonumber \\
  \!\!&=&\!\! \langle m_b |J_+ | m_i \rangle (\langle m_b |J_- | m_i \rangle)^\ast\! +\! \langle m_b |J_- | m_i \rangle (\langle m_b |J_+ | m_i \rangle)^\ast \nonumber \\
  \!\!&=&\!\! \displaystyle \sum_{m_i m_b}  2\, \mbox{Re}\left( \mathcal{A}_{+ m_i m_b} \mathcal{A}^\ast_{- m_i m_b}\right)
\end{eqnarray}

\begin{eqnarray}
T \!\!&=&\!\! \mbox{Tr}[J \sigma^i_y J^\dagger] \nonumber \\
  &=& \!\!-i \langle m_b |J_\lambda | + \rangle \langle - |J^\dagger_\lambda | m_b \rangle +i \langle m_b |J_\lambda | - \rangle \langle + |J^\dagger_\lambda | m_b \rangle \nonumber \\
  &=& \!\!-i \langle m_b |J_\lambda | + \rangle (\langle m_b |J_\lambda | - \rangle)^\ast +i \langle m_b |J_\lambda | - \rangle (\langle m_b |J^\dagger_\lambda | + \rangle)^\ast \nonumber \\
  &=& \!\!\displaystyle \sum_{\lambda m_b}  -2\, \mbox{Im}\left( \mathcal{A}_{\lambda - m_b} \mathcal{A}^\ast_{\lambda + m_b}\right)
\end{eqnarray}

\begin{eqnarray}
E \!\!&=&\!\! \mbox{Tr}[J \sigma^i_z \sigma^\gamma_z J^\dagger] \nonumber \\
  \!\!&=& \langle m_b| J_+ | m_b' \rangle \langle m_b' | \sigma_z | m_i' \rangle \langle m_i' | J_+^\dagger | m_b \rangle \nonumber \\ && \hspace{0.5cm}- \langle m_b| J_- | m_b' \rangle \langle m_b' | \sigma_z | m_i' \rangle \langle m_i' | J_-^\dagger | m_b \rangle \nonumber \\
  \!\!&=&\!\! \langle m_b| J_+ | m_b' \rangle \langle m_b' | \sigma_z | m_i' \rangle (\langle m_b | J_+ | m_i' \rangle)^\ast \nonumber \\ && \hspace{0.5cm}- \langle m_b| J_- | m_b' \rangle \langle m_b' | \sigma_z | m_i' \rangle (\langle m_b | J_- | m_i' \rangle)^\ast \nonumber \\
  \!\!&=&\!\!  \langle m_b| J_+ | + \rangle (\langle m_b | J_+ | + \rangle)^\ast - \langle m_b| J_+ | - \rangle (\langle m_b | J_+ | - \rangle)^\ast  \nonumber \\
 \!\!& &\!\! -  \langle m_b| J_- | + \rangle (\langle m_b | J_- | +  \rangle)^\ast + \langle m_b| J_- | - \rangle (\langle m_b | J_- | - \rangle)^\ast \nonumber \\
  \!\!&=&\!\! \displaystyle \sum_{m_b} \left(|\mathcal{A}_{++m_b}|^2 - |\mathcal{A}_{+-m_b}|^2 - |\mathcal{A}_{-+m_b}|^2 + |\mathcal{A}_{--m_b}|^2 \right) \nonumber \\
\end{eqnarray}

\begin{eqnarray}
F \!\!&=&\!\! \mbox{Tr}[J \sigma^i_x \sigma^\gamma_z J^\dagger] \nonumber \\
  \!\!&=&\!\! \langle m_b| J_+ | m_b' \rangle \langle m_b' | \sigma_x | m_i' \rangle \langle m_i' | J_+^\dagger | m_b \rangle \nonumber \\ && \hspace{0.5cm}- \langle m_b| J_- | m_b' \rangle \langle m_b' | \sigma_x | m_i' \rangle \langle m_i' | J_-^\dagger | m_b \rangle \nonumber \\
  \!\!&=&\!\! \langle m_b| J_+ | m_b' \rangle \langle m_b' | \sigma_x | m_i' \rangle (\langle m_b | J_+ | m_i' \rangle)^\ast \nonumber \\ && \hspace{0.5cm}- \langle m_b| J_- | m_b' \rangle \langle m_b' | \sigma_x | m_i' \rangle (\langle m_b | J_- | m_i' \rangle)^\ast \nonumber \\
  \!\!&=&\!\!  \langle m_b| J_+ | + \rangle (\langle m_b | J_+ | - \rangle)^\ast + \langle m_b| J_+ | - \rangle (\langle m_b | J_+ | + \rangle)^\ast  \nonumber \\
 \!\!& &\!\!\! -  \langle m_b| J_- | + \rangle (\langle m_b | J_- | -  \rangle)^\ast - \langle m_b| J_- | - \rangle (\langle m_b | J_- | + \rangle)^\ast \nonumber \\
  \!\!&=&\!\! \displaystyle \sum_{m_b} 2 \mbox{Re} \left( A_{++m_b} A^\ast_{+-m_b} - A_{-+ m_b} A^\ast_{--m_b} \right)
\end{eqnarray}

\begin{eqnarray}
G \!\!&=&\!\! -\mbox{Tr}[J \sigma^i_z \sigma^\gamma_y J^\dagger] \nonumber \\
  &=&\!\! i \langle m_b| J_+ | m_b' \rangle \langle m_b' | \sigma_z | m_i' \rangle \langle m_i' | J_-^\dagger | m_b \rangle \nonumber \\ && \hspace{0.5cm} -i \langle m_b| J_- | m_b' \rangle \langle m_b' | \sigma_z | m_i' \rangle \langle m_i' | J_+^\dagger | m_b \rangle \nonumber \\
  &=&\!\! i\langle m_b| J_+ | m_b' \rangle \langle m_b' | \sigma_z | m_i' \rangle (\langle m_b | J_- | m_i' \rangle)^\ast \nonumber \\ && \hspace{0.5cm}- i \langle m_b| J_- | m_b' \rangle \langle m_b' | \sigma_z | m_i' \rangle (\langle m_b | J_+ | m_i' \rangle)^\ast \nonumber \\
  &=&\!\!  i\langle m_b| J_+ | + \rangle (\langle m_b | J_- | + \rangle)^\ast -i  \langle m_b| J_+ | - \rangle (\langle m_b | J_- | - \rangle)^\ast  \nonumber \\
 & &\!\!\!\! -i  \langle m_b| J_- | + \rangle (\langle m_b | J_+ | +  \rangle)^\ast + i \langle m_b| J_- | - \rangle (\langle m_b | J_+ | - \rangle)^\ast \nonumber \\
  &=& \displaystyle \sum_{m_b} 2 \mbox{Im} \left( \mathcal{A}_{+-m_b} \mathcal{A}^\ast_{--m_b}  + \mathcal{A}_{-+m_b}\mathcal{A}^\ast_{++m_b} \right)
\end{eqnarray}

\begin{eqnarray}
H \!\!&=&\!\! -\mbox{Tr}[J \sigma^i_x \sigma^\gamma_y J^\dagger] \nonumber \\
  &=&\!\! i \langle m_b| J_+ | m_b' \rangle \langle m_b' | \sigma_x | m_i' \rangle \langle m_i' | J_-^\dagger | m_b \rangle \nonumber \\ && \hspace{0.5cm} -i \langle m_b| J_- | m_b' \rangle \langle m_b' | \sigma_x | m_i' \rangle \langle m_i' | J_+^\dagger | m_b \rangle \nonumber \\
  &=&\!\! i\langle m_b| J_+ | m_b' \rangle \langle m_b' | \sigma_x | m_i' \rangle (\langle m_b | J_- | m_i' \rangle)^\ast \nonumber \\ && \hspace{0.5cm} - i \langle m_b| J_- | m_b' \rangle \langle m_b' | \sigma_x | m_i' \rangle (\langle m_b | J_+ | m_i' \rangle)^\ast \nonumber \\
  &=&\!\!  i\langle m_b| J_+ | + \rangle (\langle m_b | J_- | - \rangle)^\ast +i  \langle m_b| J_+ | - \rangle (\langle m_b | J_- | + \rangle)^\ast  \nonumber \\
 & &\!\! -i  \langle m_b| J_- | + \rangle (\langle m_b | J_+ | -  \rangle)^\ast -i  \langle m_b| J_- | - \rangle (\langle m_b | J_+ | + \rangle)^\ast \nonumber \\
  &=&\!\!\!\! \displaystyle \sum_{m_b} 2 \mbox{Im} \left( \mathcal{A}_{--m_b} \mathcal{A}^\ast_{++m_b}  + \mathcal{A}_{-+m_b}\mathcal{A}^\ast_{+-m_b} \right)
\end{eqnarray}

\begin{eqnarray}
C_x \!\!&=&\!\! \mbox{Tr}[\sigma_x^b J \sigma^\gamma_z J^\dagger] \nonumber \\
  &=&\!\! \langle m_b | \sigma_x | m_i' \rangle  \langle m_i'| J_+ | m_i \rangle \langle m_i | J_+^\dagger | m_b \rangle \nonumber \\ && \hspace{0.5cm} - \langle m_b | \sigma_x | m_i' \rangle  \langle m_i'| J_- | m_i \rangle \langle m_i | J_-^\dagger | m_b \rangle \nonumber \\
  &=&\!\! \langle m_b | \sigma_x | m_i' \rangle  \langle m_i'| J_+ | m_i \rangle (\langle m_b | J_+ | m_i \rangle)^\ast \nonumber \\ && \hspace{0.5cm}- \langle m_b | \sigma_x | m_i' \rangle  \langle m_i'| J_- | m_i \rangle (\langle m_b | J_- | m_i \rangle)^\ast \nonumber \\
  &=& \langle - | J_+ | m_i \rangle (\langle + |J_+ | m_i \rangle )^\ast - \langle - | J_- | m_i \rangle (\langle + |J_- | m_i \rangle )^\ast \nonumber \\
  & &\!\! + \langle + | J_+ | m_i \rangle (\langle - |J_+ | m_i \rangle )^\ast - \langle + | J_- | m_i \rangle (\langle - |J_- | m_i \rangle )^\ast \nonumber \\
  &=&\!\!\! \displaystyle \sum_{m_i} 2 \mbox{Re} \left( \mathcal{A}_{+m_i-} \mathcal{A}_{+m_i+}^\ast - \mathcal{A}_{-m_i-} \mathcal{A}_{-m_i+}^\ast \right)
\end{eqnarray}

\begin{eqnarray}
C_z \!\!&=&\!\! \mbox{Tr}[\sigma_z^b J \sigma^\gamma_z J^\dagger] \nonumber \\
  &=& \langle m_b | \sigma_z | m_i' \rangle  \langle m_i'| J_+ | m_i \rangle \langle m_i | J_+^\dagger | m_b \rangle \nonumber \\ && \hspace{0.5cm} - \langle m_b | \sigma_z | m_i' \rangle  \langle m_i'| J_- | m_i \rangle \langle m_i | J_-^\dagger | m_b \rangle \nonumber \\
  &=&\!\! \langle m_b | \sigma_z | m_i' \rangle  \langle m_i'| J_+ | m_i \rangle (\langle m_b | J_+ | m_i \rangle)^\ast \nonumber \\ && \hspace{0.5cm} - \langle m_b | \sigma_z | m_i' \rangle  \langle m_i'| J_- | m_i \rangle (\langle m_b | J_- | m_i \rangle)^\ast \nonumber \\
  &=&\!\! \langle + | J_+ | m_i \rangle (\langle + |J_+ | m_i \rangle )^\ast - \langle + | J_- | m_i \rangle (\langle + |J_- | m_i \rangle )^\ast \nonumber \\
  & & - \langle - | J_+ | m_i \rangle (\langle - |J_+ | m_i \rangle )^\ast + \langle - | J_- | m_i \rangle (\langle - |J_- | m_i \rangle )^\ast \nonumber \\
  &=&\!\!\! \displaystyle \sum_{m_i} \left( |\mathcal{A}_{+m_i+}|^2 - |\mathcal{A}_{-m_i+}|^2 -|\mathcal{A}_{+m_i-}|^2 + |\mathcal{A}_{-m_i-}|^2 \right) \nonumber \\
\end{eqnarray}

\begin{eqnarray}
O_x \!\!&=&\!\! -\mbox{Tr}[\sigma_x^b J \sigma^\gamma_y J^\dagger] \nonumber \\
  &=& \!\!i\langle m_b | \sigma_x | m_i' \rangle  \langle m_i'| J_+ | m_i \rangle \langle m_i | J_-^\dagger | m_b \rangle \nonumber \\ && \hspace{0.5cm} - i \langle m_b | \sigma_x | m_i' \rangle  \langle m_i'| J_- | m_i \rangle \langle m_i | J_+^\dagger | m_b \rangle \nonumber \\
  &=& \!\!i\langle m_b | \sigma_x | m_i' \rangle  \langle m_i'| J_+ | m_i \rangle (\langle m_b | J_- | m_i \rangle)^\ast \nonumber \\ && \hspace{0.5cm} -i \langle m_b | \sigma_x | m_i' \rangle  \langle m_i'| J_- | m_i \rangle (\langle m_b | J_+ | m_i \rangle)^\ast \nonumber \\
  &=&\!\! i\langle - | J_+ | m_i \rangle (\langle + |J_- | m_i \rangle )^\ast -i \langle - | J_- | m_i \rangle (\langle + |J_+ | m_i \rangle )^\ast \nonumber \\
  & &\!\!\! + i\langle + | J_+ | m_i \rangle (\langle - |J_- | m_i \rangle )^\ast -i \langle + | J_- | m_i \rangle (\langle - |J_+ | m_i \rangle )^\ast \nonumber \\
  &=&\!\! \displaystyle \sum_{m_i} 2 \mbox{Im} \left( \mathcal{A}_{-m_i-} \mathcal{A}_{+m_i+}^\ast + \mathcal{A}_{-m_i+} \mathcal{A}_{+m_i-}^\ast \right)
\end{eqnarray}

\begin{eqnarray}
O_z \!\!&=&\!\! -\mbox{Tr}[\sigma_z^b J \sigma^\gamma_y J^\dagger] \nonumber \\
  &=&\!\! i\langle m_b | \sigma_z | m_i' \rangle  \langle m_i'| J_+ | m_i \rangle \langle m_i | J_-^\dagger | m_b \rangle \nonumber \\ && \hspace{0.5cm} - i \langle m_b | \sigma_z | m_i' \rangle  \langle m_i'| J_- | m_i \rangle \langle m_i | J_+^\dagger | m_b \rangle \nonumber \\
  &=&\!\! i\langle m_b | \sigma_z | m_i' \rangle  \langle m_i'| J_+ | m_i \rangle (\langle m_b | J_- | m_i \rangle)^\ast \nonumber \\ && \hspace{0.5cm} -i \langle m_b | \sigma_z | m_i' \rangle  \langle m_i'| J_- | m_i \rangle (\langle m_b | J_+ | m_i \rangle)^\ast \nonumber \\
  &=&\!\! i\langle + | J_+ | m_i \rangle (\langle + |J_- | m_i \rangle )^\ast -i \langle + | J_- | m_i \rangle (\langle + |J_+ | m_i \rangle )^\ast \nonumber \\
  & &\!\!\! -i \langle - | J_+ | m_i \rangle (\langle - |J_- | m_i \rangle )^\ast +i \langle - | J_- | m_i \rangle (\langle - |J_+ | m_i \rangle )^\ast \nonumber \\
  &=&\!\! \displaystyle \sum_{m_i} 2 \mbox{Im} \left( \mathcal{A}_{-m_i+} \mathcal{A}_{+m_i+}^\ast + \mathcal{A}_{+m_i-} \mathcal{A}_{-m_i-}^\ast \right)
\end{eqnarray}

\begin{eqnarray}
T_x \!\!&=&\!\! \mbox{Tr}[\sigma_x^b J \sigma_x^i J^\dagger] \nonumber \\
    &=&\!\! \langle m_b|\sigma_x| m_i \rangle \langle m_i| J_\lambda | m_i'\rangle \langle m_i' |\sigma_x| m_b'\rangle \langle m_b' |J_\lambda^\dagger | m_b\rangle \nonumber \\
    &=&\!\! \langle m_b|\sigma_x| m_i \rangle \langle m_i| J_\lambda | m_i'\rangle \langle m_i' |\sigma_x| m_b'\rangle (\langle m_b |J_\lambda| m_b'\rangle)^\ast \nonumber \\
    &=&\!\! \langle -| J_\lambda | m_i'\rangle \langle m_i' |\sigma_x| m_b'\rangle (\langle + |J_\lambda| m_b'\rangle)^\ast \nonumber \\ && \hspace{0.5cm} + \langle +| J_\lambda | m_i'\rangle \langle m_i' |\sigma_x| m_b'\rangle (\langle - |J_\lambda | m_b'\rangle)^\ast \nonumber \\
    &=&\!\! \langle -| J_\lambda | +\rangle  (\langle + |J_\lambda| -\rangle)^\ast + \langle +| J_\lambda | +\rangle (\langle - |J_\lambda| -\rangle)^\ast  \nonumber \\
    & &\!\!\!  + \langle -| J_\lambda | -\rangle (\langle + |J_\lambda | +\rangle)^\ast + \langle +| J_\lambda | -\rangle  (\langle - |J_\lambda | +\rangle)^\ast \nonumber \\
    &=&\!\! \displaystyle \sum_\lambda 2 \mbox{Re}\left( \mathcal{A}_{\lambda +-}\mathcal{A}_{\lambda -+}^\ast + \mathcal{A}_{\lambda ++}\mathcal{A}_{\lambda --}^\ast \right)
\end{eqnarray}

\begin{eqnarray}
T_z &=& \mbox{Tr}[\sigma_z^b J \sigma_x^i J^\dagger] \nonumber \\
    &=& \langle m_b|\sigma_z| m_i \rangle \langle m_i| J_\lambda | m_i'\rangle \langle m_i' |\sigma_x| m_b'\rangle \langle m_b' |J_\lambda^\dagger | m_b\rangle \nonumber \\
    &=& \langle m_b|\sigma_z| m_i \rangle \langle m_i| J_\lambda | m_i'\rangle \langle m_i' |\sigma_x| m_b'\rangle (\langle m_b |J_\lambda| m_b'\rangle)^\ast \nonumber \\
    &=& \langle +| J_\lambda | m_i'\rangle \langle m_i' |\sigma_x| m_b'\rangle (\langle + |J_\lambda| m_b'\rangle)^\ast \nonumber \\ && \hspace{0.5cm} - \langle -| J_\lambda | m_i'\rangle \langle m_i' |\sigma_x| m_b'\rangle (\langle - |J_\lambda | m_b'\rangle)^\ast \nonumber \\
    &=& \langle +| J_\lambda | +\rangle  (\langle + |J_\lambda| -\rangle)^\ast - \langle -| J_\lambda | +\rangle (\langle - |J_\lambda| -\rangle)^\ast  \nonumber \\
    & & + \langle +| J_\lambda | -\rangle (\langle + |J_\lambda | +\rangle)^\ast - \langle -| J_\lambda | -\rangle  (\langle - |J_\lambda | +\rangle)^\ast \nonumber \\
    &=& \displaystyle \sum_\lambda 2 \mbox{Re}\left( \mathcal{A}_{\lambda ++}\mathcal{A}_{\lambda -+}^\ast - \mathcal{A}_{\lambda +-}\mathcal{A}_{\lambda --}^\ast \right)
\end{eqnarray}

\begin{eqnarray}
L_x &=& \mbox{Tr}[\sigma_x^b J \sigma_z^i J^\dagger] \nonumber \\
    &=& \langle m_b|\sigma_x| m_i \rangle \langle m_i| J_\lambda | m_i'\rangle \langle m_i' |\sigma_z| m_b'\rangle \langle m_b' |J_\lambda^\dagger | m_b\rangle \nonumber \\
    &=& \langle m_b|\sigma_x| m_i \rangle \langle m_i| J_\lambda | m_i'\rangle \langle m_i' |\sigma_z| m_b'\rangle (\langle m_b |J_\lambda| m_b'\rangle)^\ast \nonumber \\
    &=& \langle -| J_\lambda | m_i'\rangle \langle m_i' |\sigma_z| m_b'\rangle (\langle + |J_\lambda| m_b'\rangle)^\ast + \langle \nonumber \\ && \hspace{0.5cm} +| J_\lambda | m_i'\rangle \langle m_i' |\sigma_z| m_b'\rangle (\langle - |J_\lambda | m_b'\rangle)^\ast \nonumber \\
    &=& \langle -| J_\lambda | +\rangle  (\langle + |J_\lambda| +\rangle)^\ast + \langle +| J_\lambda | +\rangle (\langle - |J_\lambda| +\rangle)^\ast  \nonumber \\
    & &  - \langle -| J_\lambda | -\rangle (\langle + |J_\lambda | -\rangle)^\ast - \langle +| J_\lambda | -\rangle  (\langle - |J_\lambda | -\rangle)^\ast \nonumber \\
    &=& \displaystyle \sum_\lambda 2 \mbox{Re}\left( \mathcal{A}_{\lambda +-}\mathcal{A}_{\lambda ++}^\ast - \mathcal{A}_{\lambda --}\mathcal{A}_{\lambda -+}^\ast \right)
\end{eqnarray}

\begin{eqnarray}
L_z &=& \mbox{Tr}[\sigma_z^b J \sigma_z^i J^\dagger] \nonumber \\
    &=& \langle m_b|\sigma_z| m_i \rangle \langle m_i| J_\lambda | m_i'\rangle \langle m_i' |\sigma_z| m_b'\rangle \langle m_b' |J_\lambda^\dagger | m_b\rangle \nonumber \\
    &=& \langle m_b|\sigma_z| m_i \rangle \langle m_i| J_\lambda | m_i'\rangle \langle m_i' |\sigma_z| m_b'\rangle (\langle m_b |J_\lambda| m_b'\rangle)^\ast \nonumber \\
    &=& \langle +| J_\lambda | m_i'\rangle \langle m_i' |\sigma_z| m_b'\rangle (\langle + |J_\lambda| m_b'\rangle)^\ast \nonumber \\ && \hspace{0.5cm} - \langle -| J_\lambda | m_i'\rangle \langle m_i' |\sigma_z| m_b'\rangle (\langle - |J_\lambda | m_b'\rangle)^\ast \nonumber \\
    &=& \langle +| J_\lambda | +\rangle  (\langle + |J_\lambda| +\rangle)^\ast - \langle -| J_\lambda | +\rangle (\langle - |J_\lambda| +\rangle)^\ast  \nonumber \\
    & &  - \langle +| J_\lambda | -\rangle (\langle + |J_\lambda | -\rangle)^\ast + \langle -| J_\lambda | -\rangle  (\langle - |J_\lambda | -\rangle)^\ast \nonumber \\
    &=& \displaystyle \sum_\lambda \left( |\mathcal{A}_{\lambda++}|^2 -|\mathcal{A}_{\lambda+-}|^2 - |\mathcal{A}_{\lambda-+}|^2 + |\mathcal{A}_{\lambda--}|^2\right) \nonumber \\
\end{eqnarray}

\subsection{Expressions in terms of $L_i$ amplitudes}

A summary of the expressions for the 16 observables in terms of the $L_i$ amplitudes as given below: 
\begin{subequations}
\label{eqn:long_basis_obs}
\begin{eqnarray}
\sigma_0/2 &=& (|L_1|^2 + |L_2|^2 + |L_3|^2 + |L_4|^2) \\
P &=& - 2 \mbox{Im} (L_1 L^\ast_2 + L_4 L^\ast_3) \\
\Sigma &=& 2 \mbox{Re} (L_1 L_3^\ast -L_4 L_2^\ast) \\
T &=& 2 \mbox{Im} (L_1 L^\ast_4 + L_2 L_3^\ast) \\
E &=& (|L_1|^2 + |L_2|^2 -|L_3|^2 -|L_4|^2)\\
F &=& 2 \mbox{Re} (L_1 L_4^\ast + L_2 L_3^\ast)\\
G &=& -2 \mbox{Im} (L_1 L_3^\ast -L_2 L_4^\ast)\\
H &=& 2 \mbox{Im} (L_1 L_2^\ast + L_3 L_4^\ast ) \\
C_x &=& 2 \mbox{Re} (L_1 L_2^\ast + L_3 L_4^\ast)\\
C_z &=& (|L_1|^2 - |L_2|^2 -|L_3|^2 + |L_4|^2)\\
O_x &=& 2 \mbox{Im} (L_1 L_4^\ast - L_2 L_3^\ast)\\
O_z &=& -2 \mbox{Im} (L_1 L_3^\ast + L_2 L_4^\ast)\\
T_x &=& 2 \mbox{Re} (L_1 L_3^\ast + L_2 L_4^\ast)\\
T_z &=& 2 \mbox{Re} (L_1 L_4^\ast - L_2 L_3^\ast)\\
L_x &=& 2 \mbox{Re} (L_1 L_2^\ast - L_3 L_4^\ast) \\
L_z &=& (|L_1|^2 -|L_2|^2 +|L_3|^2 - |L_4|^2), 
\end{eqnarray}
\end{subequations}
where it is understood that all the 15 polarization observables are to be normalized by $\sigma_0/2$. 

\section{The consistency relations}

It is well known that the fifteen polarization observables occurring as bilinears in Eq.~\ref{eqn:long_basis_obs} can be connected by various identities. These are also called constraint equations, because they interconnect and place restrictions on the physical values these observables can take. Simply put, these equations are nothing but identities in the four independent amplitudes $L_i$. Chiang and Tabakin~\cite{ct} have showed that these identities can be derived in a more sophisticated fashion by considering the complex space spanned by the four amplitudes. The observables can then be expanded in terms of the sixteen $4 \times 4$ Dirac gamma matrix bilinears $\{1,\gamma^\mu,\sigma^{\mu \nu}, \gamma^{\mu \nu\rho}, \gamma^{\mu \nu \rho \sigma} \}$ and the constraint relations emerge from the various Fierz identities connecting products of the Dirac bilinears. We list the set of relations (Eqs.~L.0-S.r) that we find to be valid (we maintain the equation-numbering as in Chiang-Tabakin~\cite{ct}). 

These relations have been numerically verified by assigning random values to the four complex $L_i$ amplitudes and calculating the polarizations employing Eqs.~\ref{eqn:long_basis_obs}a-p. The relations consisting of only squares of the observables (Eqs.~L.0 and S.bt-S.r) have no sign ambiguities. However, the signs in the remaining set of relations depend on the conventions adopted while defining the polarizations.

There appears to be some disagreement between different groups in the sign conventions for the polarizations, most likely arising from differences in the physics motivation. For example, in the CLAS $C_x/C_z$ measurements for $K^+\Lambda$ photoproduction~\cite{bradford}, it was found that $C_z \to +1$ at $\vartheta_{\mbox{\scriptsize c.m.}}^{K^+} \to 0$ and $C_z$ was seen as the {\em spin}-transfer from a right-handed circularly polarized photon to the recoiling baryon. Other groups~\cite{said_maid} prefer to have  $C_z \to +1$ at $\vartheta_{\mbox{\scriptsize c.m.}}^{\Lambda} \to 0$, with the interpretation that $C_z$ is the transfer of {\em helicity} from a right-handed circularly polarized photon to the $\Lambda$. Indeed, hadron-helicity-conservation is a feature of perturbative QCD at high enough energies~\cite{hhc}. Whatever be the choice of convention, the important issue is that the intensity profile the experimentalist uses must match with the asymmetry definitions that give the amplitude-level expressions. This point was detailed in Sec.~\ref{sec:experimental_profiles}.  

\begin{align*}
1 &= \{ \Sigma^2 + T^2 + P^2 + E^2 + G^2 + F^2 + H^2 \notag\\
  & + O_x^2 + O_z^2 + C_x^2 + C_z^2 + L_x^2 + L_z^2 + T_x^2 + T_z^2\} /3 \tag{L.0}\\
\Sigma &= TP + T_x L_z - T_z L_x \tag{L.tr}\\
T  &= \Sigma P - (C_x O_z - C_z O_x) \tag{L.br}\\
P &= \Sigma T + GF - E H \tag{L.bt}\\
G &= PF + O_x L_x + O_z L_z \tag{L.1}\\
H &=  -P E + O_x T_x + O_z T_z \tag{L.2}\\
E &= -P H  + C_x L_x + C_z L_z \tag{L.3}\\
F &= PG + C_x T_x + C_z T_z \tag{L.4}\\
O_x &=  T C_z + G L_x  +H T_x \tag{L.5}\\
O_z &= -T C_x + G L_z + HT_z \tag{L.6}\\
C_x &= -TO_z + EL_x + F T_x \tag{L.7}\\
C_z &=  T O_x +E L_z + F T_z\tag{L.8}\\
T_x &= \Sigma L_z + H O_x + F C_x  \tag{L.9}\\
T_z &= -\Sigma L_x + H O_z + F C_z \tag{L.10}\\
L_x &= -\Sigma T_z + G O_x + E C_x \tag{L.11}\\
L_z &= \Sigma T_x + G O_z + E C_z \tag{L.12}\\
0 &= C_x O_x + C_z O_z - E G - F H\tag{Q.b}\\
0 &= GH + EF -L_xT_x - L_zT_z \tag{Q.t}\\
0 &= C_x C_z + O_x O_z -L_x L_z - T_x T_z \tag{Q.r}\\
0 &= -\Sigma G + TF + O_z T_x - O_x T_z  \tag{Q.bt.1}\\
0 &= -\Sigma H - TE -  O_z L_x + O_x L_z\tag{Q.bt.2}\\
0 &= \Sigma E + T H - C_z T_x + C_x T_z\tag{Q.bt.3}\\
0 &= - \Sigma F + TG - C_z L_x + C_x L_z \tag{Q.bt.4}\\
0 &= -\Sigma O_x + P C_z - GT_z + H L_z  \tag{Q.br.1}\\
0 &= -\Sigma O_z - P C_x + G T_x - H L_x  \tag{Q.br.2}\\
0 &= -\Sigma C_x - P O_z - ET_z + F L_z \tag{Q.br.3}\\
0 &= -\Sigma C_z + P O_x + E T_x - F L_x\tag{Q.br.4}\\
0 &= TT_x - PL_z - H C_z + F O_z \tag{Q.tr.1}\\
0 &= T T_z + P L_x + H C_x - F O_x \tag{Q.tr.2}\\
\end{align*}

\begin{align*}
0 &= T L_x + PT_z - G C_z + E O_z \tag{Q.tr.3}\\
0 &= TL_z - P T_x + G C_x - E O_x \tag{Q.tr.4}\\
1 &= G^2 + H^2 +E^2 +F^2 +\Sigma^2 + T^2 - P^2 \tag{S.bt}\\
1 &= O_x^2 + O_z^2 + C_x^2 + C_z^2 + \Sigma^2 - T^2 + P^2 \tag{S.br}\\
1 &= T_x^2 + T_z^2 + L_x^2 + L_z^2 - \Sigma^2 + T^2 + P^2 \tag{S.tr}\\
0 &= G^2 + H^2 -E^2 -F^2 - O_x^2 - O_z^2 + C_x^2 + C_z^2 \tag{S.b}\\
0 &= G^2 - H^2 +E^2 -F^2 + T_x^2 + T_z^2 -L_x^2 -L_z^2 \tag{S.t}\\
0 &= O_x^2 - O_z^2 + C_x^2 - C_z^2 - T_x^2 + T_z^2 - L_x^2 + L_z^2 \tag{S.r}
\end{align*}

\section{Summary and outlook}

We provide a detailed and self-contained description of the intensity profiles and amplitude-level expressions for the 15 polarization observables in pseudo-scalar meson photoproduction. Our calculations are based on the density matrix approach of Fasano, Tabakin, and Saghai and our spin amplitudes have a universal spin-projection direction along the incident beam direction for all particles. We have also stressed the preservation of consistency between the sign-conventions of experiment and theoretical amplitude-level expressions. The current work is geared towards performing a mass-independent partial-wave analysis on the recently published CLAS data~\cite{bradford-dcs, bradford, prc_klam, prc_ksig} and forthcoming results from JLab~\cite{g8,frost}.

\section{Acknowledgements}
The authors thank Professor Reinhard Schumacher for his helpful comments during the preparation of this article. This work was supported in part by the US Department of Energy under Grant No.~DE-FG02-87ER40315. D.~Ireland also acknowledges the support of the United Kingdom's Science and Technology Council.

\end{document}